\documentclass[twocolumn,aps,amssymb,prl,showpacs]{revtex4}
\usepackage{latexsym}
\usepackage{graphicx}
\usepackage{color}

\begin{document}



\title{ Observation of thermally activated glassiness and memory dip in a-NbSi insulating thin films}
\author{J. Delahaye}
\affiliation{Institut N\'eel, CNRS and Universit\'e Joseph Fourier, BP 166, 38042 Grenoble, France}
 \author{T. Grenet}
 \affiliation{Institut N\'eel, CNRS and Universit\'e Joseph Fourier, BP 166, 38042 Grenoble, France}
 \author{C.A. Marrache-Kikuchi}
 \affiliation{CSNSM, Universit\'e Paris-Sud, Orsay, F-91405, France}
 \author{A.A. Drillien}
 \affiliation{CSNSM, Universit\'e Paris-Sud, Orsay, F-91405, France}
  \author{L. Berg\'e}
 \affiliation{CSNSM, Universit\'e Paris-Sud, Orsay, F-91405, France}

\date{\today} 

\begin{abstract}
We present electrical conductance measurements on amorphous NbSi insulating thin films. These films display out-of equilibrium electronic features that are markedly different from what has been reported so far in disordered insulators. Like in the most studied systems (indium oxide and granular Al films), a slow relaxation of the conductance is observed after a quench to liquid helium temperature which gives rise to the growth of a memory dip in MOSFET devices. But unlike in these systems, this memory dip and the related conductance relaxations are still visible up to room temperature, with clear signatures of a temperature dependent dynamics.
\end{abstract}

 \bigskip
 \maketitle


Since the pioneering works of Zvi Ovadyahu and coworkers 20 years ago \cite{OvadyahuPRB91}, the number of disordered insulating systems in which glassy conductance relaxations have been reported is slowly increasing with time \cite{ElectronGlassBook}. It gathers now amorphous and micro-crystalline indium oxide films \cite{OvadyahuPRB91}, granular aluminium films \cite{GrenetEPJB03}, Be films \cite{OvadyahuPRB10}, ultrathin and discontinuous films of metals \cite{GoldmanPRB98,FrydmanEPL12} and thallium oxide films \cite{OvadyahuPRB13}. Until very recently, the glassy features like the memory dip and its activationless logarithmic slow relaxation were believed to be of universal character. However a new behaviour was evidenced two years ago in discontinuous metal films \cite{FrydmanEPL12}: the conductance relaxations were found to be strongly suppressed below a well-defined temperature T*, whereas nothing similar was seen in indium oxide \cite{OvadyahuEPL98,OvadyahuPRL07} and granular Al films \cite{GrenetEPJB07}, the two most extensively studied systems so far.

These glassy features remain for a large part unexplained but they might be the experimental signature of an electron glass \cite{ElectronGlassTheory,ElectronGlassRecent,ElectronGlassBook}. This hypothesis is supported by the fact that all systems in which out-of-equilibrium effects are observed have a large charge carrier density compared to standard doped semiconductors close to the metal-insulator transition \cite{OvadyahuPRL98,OvadyahuPRB13}. In indium oxide films, the charge carrier density was indeed found to influence the gate voltage width of the memory dip and the conductance dynamics itself \cite{OvadyahuPRL98}, even if aspects of this last result were questioned recently \cite{GrenetPRB12}.

In order to make some progress towards the understanding of these phenomena, it is of crucial importance to identify among the observed properties what is universal and what is specific to each system. In this respect, the exploration of new systems is an incomparable source of information. We present here the first investigation of out of equilibrium phenomena  in amorphous (a-) insulating NbSi thin films. We show that this system also displays slow conductance relaxations after a quench from room to liquid helium temperature, as well as gate voltage ($V_g$) history memory. However the characteristics of the memory dip as well as the effects of temperature are different from all known systems and strongly indicate a thermal activation of the dynamics \cite{OvadyahuPRL07,OvadyahuEPL98,GrenetEPJB07}.


Our NbSi films were obtained by co-deposition of Nb and Si at room temperature and under ultrahigh vacuum (typically a few $10^{-8}$ mbar) \cite{CraustePRB13}. Samples for conductance relaxation measurements (see below) were deposited on sapphire substrates coated with a 25 nm thick SiO underlayer designed to smooth the substrate and were protected from oxidation by a 25 nm thick SiO overlayer. Samples for electrical field effect measurements were deposited on Si++ wafers (the gate) coated with 100 nm of thermally grown $SiO_2$ (the gate insulator). These were subsequently covered with a 12.5 nm thick SiO overlayer. Previous studies have shown that such films are continuous down to a thickness of 2 nm and that they are amorphous and homogeneous down to the nanometre scale \cite{CraustePRB13}. The SiO under- and over- layers were found to play no significant role in the electrical glassy behaviour described below \cite{NoteSiO}.

Electrical measurements were done either in two or four contacts configurations. Voltage or current bias was limited to low enough values in order to stay in the ohmic regime. The resistance of the films has an exponential-like divergence at low temperature of the form $R \propto \exp(T_0/T)^\alpha$, with $0.5<\alpha<1$.

\begin{figure}
    \includegraphics[width=8cm]{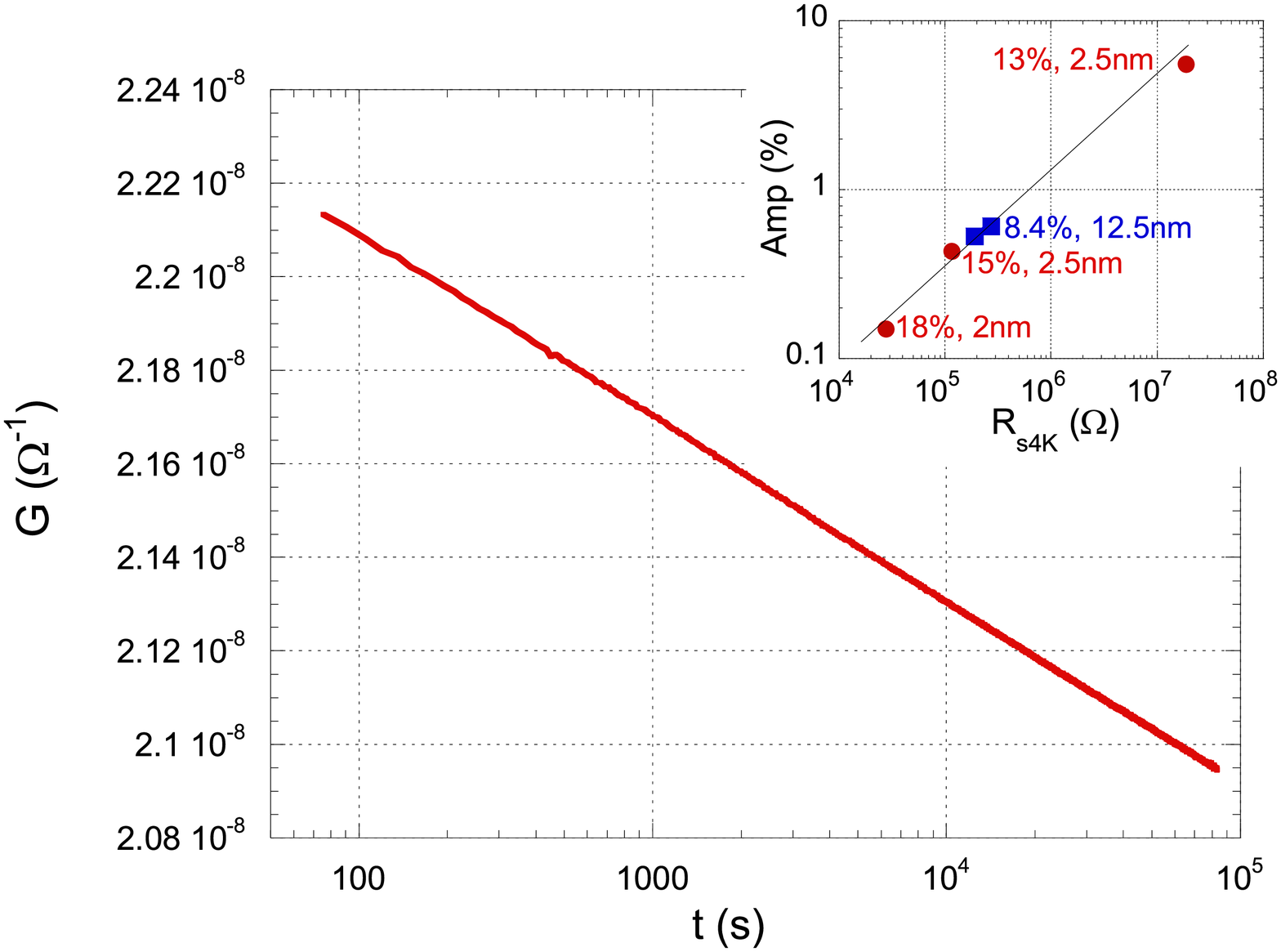}
    \caption{Conductance as a function of time after a cooling down from 300K to 4.2K for an a-NbSi film. Film parameters: thickness = 2.5nm; Nb content = 13\% and $R_s(4.2K)=20M\Omega$. Insert: amplitude of conductance relaxations for different a-NbSi films (see the text for details). For each film, the Nb content and the thickness are indicated. The straight line is a guide for the eyes.}\label{Figure1}
    \end{figure}

We have first measured the conductance variations of the films deposited on sapphire after a rapid (about 10mn) cooling down from room temperature to liquid helium. Our experimental set-up was already described in details elsewhere \cite{DelahayePRL11}. A typical result is shown in Figure \ref{Figure1} for a 2.5nm thick $a-Nb_{0.13}Si_{0.87}$ film. Once at 4.2K, the conductance is found to decrease as a logarithm of the time elapsed since the cooling down, with no signs of saturation even after several days of measurements. If we define the relaxation amplitude as the conductance change between 100s and $10^5s$, it reaches 5\% in this sample (sheet resistance $R_s$ at 4K of about $20M\Omega$). Logarithmic conductance relaxations were found in all the samples we have measured. The relative amplitude of the relaxation increases with $R_s$ but no significant difference was observed between 12.5nm and 2.5nm thick films of similar $R_s$ (see the insert of Figure \ref{Figure1}). This is qualitatively similar to what is seen in indium oxide
\cite{OvadyahuPRB02,OvadyahuPRB03,OvadyahuPRB06} and granular Al thin films \cite{GrenetEPJB07,DelahayePRL11}.

\begin{figure}
    \includegraphics[width=8cm]{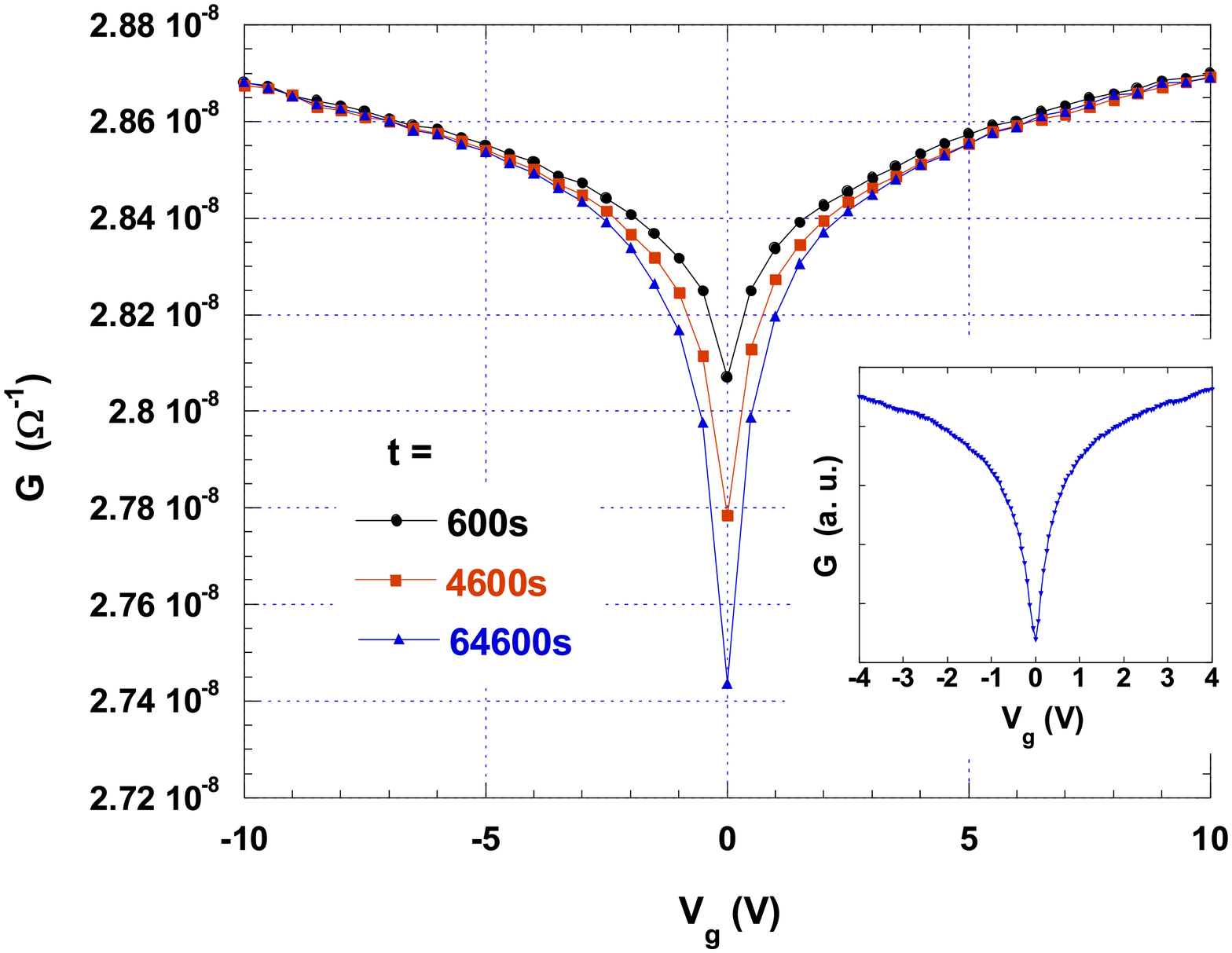}
    \caption{$G(V_g)$ curves measured a time t after a cooling down from 300K to 4.2K under $V_{geq}=0V$ (see the text for the details).  Insert: zoom on a smaller voltage range. Film parameters: thickness = 2.5nm; Nb content = 13\% and $R_s(4.2K)=100M\Omega$.}\label{Figure2}
    \end{figure}

A set of $a-Nb_{0.13}Si_{0.87}$ films 2.5nm thick with $R_s(4K)\simeq 100M\Omega$ was also deposited on a Si++/$SiO_2$ substrate in order to perform field effect measurements. The films were first cooled down from 300K to 4.2K under a gate voltage $V_g$ of 0V. Once at 4.2K, $V_g$ sweeps from -30V to +30V were repeated at constant time intervals while the $V_g$ value was maintained at 0V between the sweeps. The $V_g$ value maintained between the sweeps is called the “equilibrium” gate voltage and is noted $V_{geq}$. Typical $G(V_g)$ curves are shown in Figure \ref{Figure2}. A conductance dip or memory dip of a few \% centred on $V_{geq}$ is clearly visible and its amplitude increases as a function of time. Once again, this is qualitatively similar to what is seen in indium oxide \cite{OvadyahuPRB02} and granular Al thin films \cite{GrenetEPJB07,DelahayePRL11}.

\begin{figure}
    \includegraphics[width=8cm]{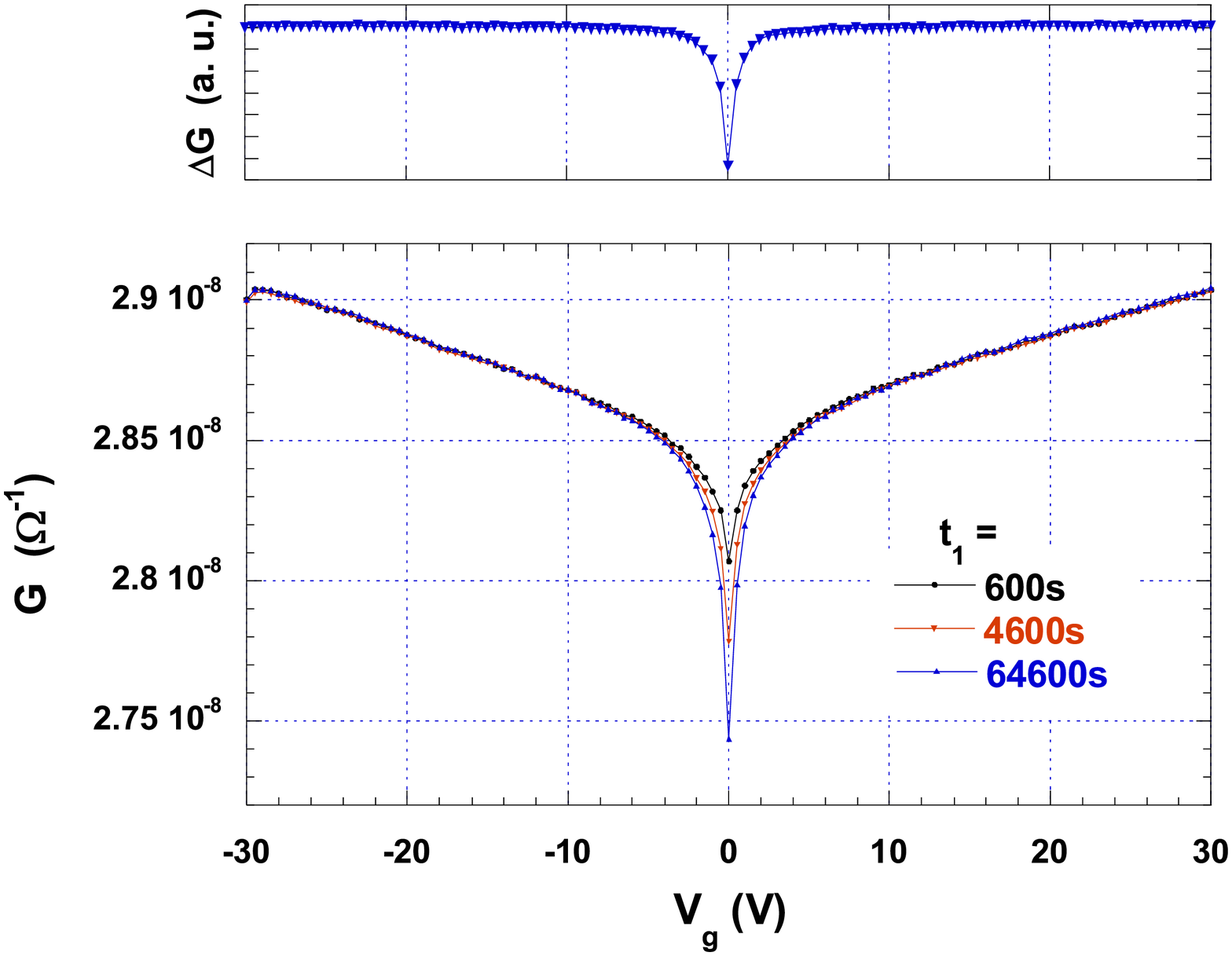}
    \caption{Same data as in Figure \ref{Figure2} but on a larger $V_g$ scale. Upper graph: difference in arbitrary units between two $G(V_g)$ curves: one measured a few minutes after the cooling down and the other one about 20h later.}\label{Figure3}
    \end{figure}

But when the same $G(V_g)$ curves are plotted on a larger $V_g$ scale (see Figure \ref{Figure3}), another feature becomes noteworthy. We observe no saturation of the conductance increase when $V_g$ is scanned away from $V_{geq}$ even up to 30V. By contrast, a saturation to a constant value or to a small normal field effect is present in all other systems studied so far. Thus the memory dip, which reflects the sample memory of its $V_g$ history, is unusually wide in NbSi films compared to granular Al films \cite{GrenetEPJB07} and highly doped indium oxide films \cite{OvadyahuPRL98}, systems exhibiting the broadest memory dips. A $V_g$ value of 30V over a $SiO_2$ layer of 100nm corresponds to a surface charge density of $\simeq6 \times 10^{12}e/cm^2$. In granular Al films the conductance dip at 4.2K is limited to changes of the surface charge densities at least 6 times smaller ($V_g$ values of about 5V in the $V_g$ scale of Figures \ref{Figure2} and \ref{Figure3}). But if the dip is very wide, only part of it is changing after the cooling down. As highlighted by the upper graph of Figure \ref{Figure3}, the changes after a cool down at 4.2K are limited to 10V around $V_{geq}$ while the rest of the dip remains unchanged, as if it was frozen.

\begin{figure}
    \includegraphics[width=8cm]{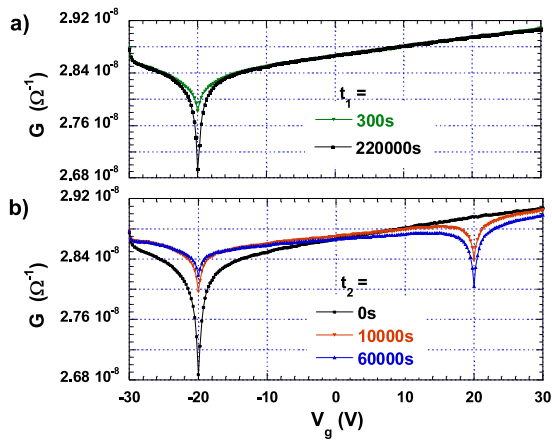}
    \caption{(a) $G(V_g)$ curves measured as a function of the time $t_1$ elapsed since the cooling down from 300K to 4.2K ($V_{geq} = -20V$). (b) $G(V_g)$ curves measured a time $t_2$ after a $V_{geq}$ change from -20V to +20V.}\label{Figure4}
    \end{figure}

A cooling down from 300K to 4.2K under a different $V_{geq}$ (-20V in Figure \ref{Figure4}a) results in a broad dip centred on this new value. Thus both the ``frozen'' and the growing parts of the dip reflect the $V_g$ memory of the sample. If, at low T, we change $V_{geq}$ from $V_{geq1} = -20V$ to $V_{geq2} = 20V$ (Figure \ref{Figure4}b), a new dip forms at $V_{geq2}$ while the old one centred on $V_{geq1}$ is slowly erased. It is noteworthy that the $V_g$ induced relaxation is not limited within 10V around $V_{geq1}$ and $V_{geq2}$, but affects the whole $V_g$ range. In other words, the broad ``frozen'' part is set to relax by a $V_{geq}$ change.

\begin{figure}
    \includegraphics[width=8cm]{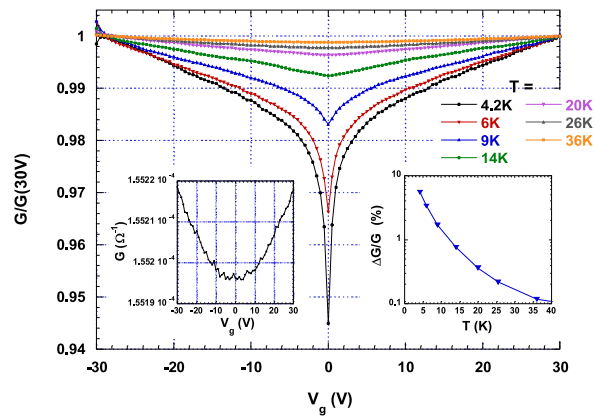}
    \caption{Conductance dip measured at different T and normalized at $V_g=30V$. The sample was first cooled down from 300K to 4.2K under $V_{geq} = 0V$ and let at 4.2K for a few days before any T changes. The temperature was then increased, by steps, from 4.2 to 36K and kept constant during 1h before a $G(V_g)$ scan (no significant changes in the amplitude are observed beyond this time scale). Left-side insert : $G(V_g)$ scan at 300K (a linear background has been subtracted)\cite{Dip300K}. Right-side insert: relative conductance difference between $V_g = 30V$ and $0V$ as a function of T.}\label{Figure5}
    \end{figure}

In Figure \ref{Figure5}, normalized memory dips are shown at different temperatures from 4.2K up to 36K. They have been measured after a few days at 4.2K under $V_{geq} = 0V$. Since the limits of the dips are out of our available $V_g$ window, it is not possible to define precisely their amplitudes and their widths. However, the relative difference between the conductance measured at $V_g = 30V$ and $0V$ gets smaller under a temperature increase. Interestingly enough, this decrease of the dip amplitude with T appears to be weaker than what was seen in granular Al and indium oxide films \cite{OvadyahuEPL98,GrenetEPJB07}. A trace of the conductance dip is even visible at room temperature (see the left-side insert of Figure \ref{Figure5}) while $R_s$ is of only $25k\Omega$. The conductance difference between $V_g = 30V$ and $0V$ is then of the order of 0.01\%. A room temperature dip was never observed in previously studied systems, except recently in discontinuous Au films made at room temperature \cite{FrydmanPrivateCom}. The shape of the dip is also temperature dependent: it gets rounder at higher T.

We come now to the T dependence of the conductance relaxation dynamics. Such dependence cannot be revealed simply by measuring the conductance relaxations after a $V_g$ change or a quench at different T. The relaxations are logarithmic in time (see Figure \ref{Figure1}) and they thus contain no characteristic times by themselves. More complex protocols have to be used and to this respect, the so called ``two dip'' protocol turned out to be a powerful tool \cite{OvadyahuPRL97,OvadyahuPRL98,OvadyahuPRB02}. In one version of this protocol, a new conductance dip is formed by fixing $V_{geq}$ to a never explored value $V_{geq1}$ during a time $t_w$ (the ``writing'' step). Then, $V_{geq}$ is changed to a different value $V_{geq2}$ and the erasure of the $V_{geq1}$ dip amplitude is measured as a function of time (the ``erasure'' step). It was found in indium oxide and granular Al films that the $V_{geq1}$ dip erasure scales with $t/t_w$ \cite{OvadyahuPRL00,OvadyahuPRB02,GrenetEPJB07} and that the characteristic erasure time is equal to $t_w$ \cite{NoteTime}, meaning that it takes a typical time $t_w$ to erase a dip formed during a time $t_w$. This finding can be simply explained by assuming that the writing and the erasure of the memory dip result from the same modes switching back and forth with unchanged characteristic relaxation times under $V_{geq}$ changes \cite{GrenetEPJB07,AmirPRL09}.

\begin{figure}
    \includegraphics[width=8cm]{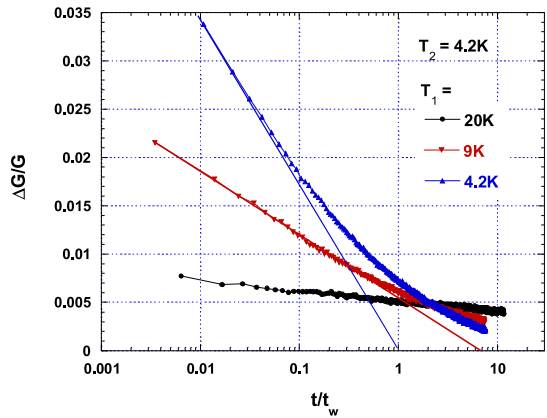}
    \caption{Amplitude of a dip formed during $t_w=20000s$ under $V_{geq1}=20V$ at $T_1=4.2K$, 9K and 20K and measured at $T_2=4.2K$ after a $V_{geq}$ change from 20V to 0V. The $V_{geq}$ change was done just before the cooling down from $T_1$ to $T_2$, so that the writing step was always made at a well defined temperature. The intercept between the straight lines and the abscissa axis defines the characteristic erasing time \cite{NoteTime}.}\label{Figure6}
    \end{figure}

In order to reveal a T dependence of the relaxation dynamics, the writing and the erasure of the $V_{geq1}$ dip have to be done at two different T \cite{GrenetEPJB07}. In Figure 6, the erasures at $T_2 = 4.2K$ of memory dips formed during $t_w=20000s$ at different $T_1 \geq 4.2K$ are compared. When $T_1=T_2=4.2K$, we get a ``trivial'' result, i.e. it takes a characteristic time of about $t_w$ to erase a dip formed during $t_w$. But when the writing is done at $T_1 > T_2 = 4.2K$, the situation is different. Beyond the change of the $V_{geq1}$ dip amplitude at short times, the characteristic erasure time, if any, is now larger than $t_w$ (the erasure curve corresponding to $T_1=20K$ is indeed almost constant). Assuming that the same modes are involved in the writing and the erasure of the dip, these results indicate a slow down of the dynamics under cooling: the modes are slower at $T_2=4.2K$ than at $T_1>T_2$. Similar measurements performed in granular Al films below $T\simeq 10K$ have revealed no T dependence of the characteristic erasure time: it is always equal to $t_w$, whatever $T_1$ and $T_2$ \cite{GrenetEPJB07}. The results of Figure \ref{Figure6} also suggest that the memory of any $V_g$ visited at various T remains printed at lower T. Further investigations have confirmed that the broad ``frozen'' part of the dip shown in Figures \ref{Figure3} and \ref{Figure4} reflects the accumulation of memory during the cooling down.

Extensive studies performed in indium oxide and granular Al films have shown that these two systems display very similar glassy features: logarithmic conductance relaxation after a quench and memory dip in MOSFET devices whose dynamics is essentially T independent \cite{OvadyahuPRL07,GrenetEPJB07}. Since these features were observed in micro-crystalline, amorphous and granular samples, it was tempting to believe that they should be rather universal. A first breach in this universal picture was recently opened by conductance measurements on discontinuous films of metals \cite{FrydmanEPL12}. In such films, a memory dip is present but its dynamics is strongly T dependent: the dip is frozen below a temperature T* which is determined by the highest T experienced by the sample, suggesting that a well defined activation energy dominates the dynamics of the system. Since discontinuous films of metals have a very specific (maze) microstructure, it may not be surprising to observe also a specific T dynamics of their memory dip.

Our results demonstrate that non-universal behaviors are not limited to ``exotic'' microstructures. In a-NbSi films, the dynamics of the memory dip is also found to be T dependent. But in contrary to discontinuous films of metals, the dip responds to $V_{geq}$ changes down to 4.2K, suggesting that its dynamics is governed by a large distribution of activation energies. Why amorphous NbSi films do not behave like amorphous indium oxide films is a challenging question. According to specific heat measurements \cite{MarnierosPhysicaB99}, the charge carrier densities of (weakly) insulating a-NbSi films lie in the range of $10^{22}-10^{23}cm^{-3}$. Such values are similar to the largest charge carrier densities reported in indium oxide films \cite{OvadyahuPRL98}. A more promising issue might be to consider the ionic character of the two systems. Indium oxide and NbSi are expected to be respectively ionic and covalent alloys, a difference which should strongly influence the potentiel lanscape and the statistics of the localized states.


In summary, our electrical measurements on a-NbSi films have revealed out-of-equilibrium and glassy features. A slow relaxation of the conductance is seen after a cool down of the films from room temperature to liquid helium and a conductance dip centred on the equilibrium gate voltage is observed in electrical field effect measurements. Compared to granular Al and indium oxide thin films, the dip is very wide, robust under a temperature increase and its dynamics seem to be strongly temperature dependent. Our findings strengthen the fact that important differences exist among the glassy features of disordered insulating systems. They are much more diverse that what was believed until recently and cannot be reduced to the prototypical case of indium oxide films. Theoretical models should also not be restrained to T independent tunneling dynamics. Where these differences come from is a crucial but so far unsolved question. In a-NbSi films, the electrical resistance can be tuned by changing the Nb content, the thickness or by annealing the films up to different temperatures \cite{CraustePRB13}. By testing the influence of these parameters on the electrical glassy properties, we have the unique opportunity to better understand their physical origin.

\begin{acknowledgments}
Discussions with O. Crauste and A. Frydman are gratefully acknowledged.
\end{acknowledgments}




\begin{thebibliography}{99}
\bibitem{NoteSiO} The conductance relaxations of two 12.5nm thick NbSi films deposited on sapphire with and without the SiO underlayer were measured after a quench from room T to 4.2K. We found the same logarithmic decrease with the sample amplitude in both cases. Moreover, field effect measurements on 12.5nm thick films without any SiO overlayer give similar glassy features to what is described below on 2.5nm thick film with a SiO overlayer.
\bibitem{Dip300K} The conductance minimum can be moved at 300K to new $V_g$ values after long stays under different $V_{geq}$.
\bibitem{NoteTime} The characteric erasure time is defined as the intercept between the extrapolated short time logarithmic dependence and the abscissa axis.
\end{thebibliography}
\end{document}